\documentclass [aip,jmp, amsmath,amssymb,reprint]{revtex4-1}

\usepackage{amssymb}
\usepackage{amsbsy}
\usepackage{amsmath}
\usepackage{amsfonts}
\usepackage{mathrsfs}
\usepackage{latexsym}
\usepackage[english]{babel}
\usepackage{url}
\usepackage[]{graphicx}
\usepackage{subfigure}
\usepackage{hyperref}

\begin{document}


\title{Circular single domains in hemispherical Permalloy nanoclusters}

\author{Clodoaldo I. L. de Araujo}
 \altaffiliation{Departamento de F\'{i}sica, Universidade Federal de Vi\c{c}osa, 36570-900, Vi\c{c}osa-MG, Brazil.}
  \email{dearaujo@ufv.br}
 
 \author{Jakson M. Fonseca}%
 \affiliation{Departamento de F\'{i}sica, Universidade Federal de Vi\c{c}osa, 36570-900, Vi\c{c}osa-MG, Brazil.}%

\author{Jo\~ao P. Sinnecker}
\affiliation{Centro Brasileiro de Pesquisas F\'{i}sicas, 22290-180, Rio de Janeiro-RJ, Brazil}%

\author{Rafael G. Delatorre}
\affiliation{Departamento de F\'{i}sica, Universidade Federal de Santa Catarina, 88040-900, Florian\'{o}polis-SC, Brazil}

  \author{Nicolas Garcia}
  \affiliation{Departamento de F\'{i}sica, Universidade Federal de Santa Catarina, 88040-900, Florian\'{o}polis-SC, Brazil}

  \author{Andr\'{e} A. Pasa}
  \affiliation{Departamento de F\'{i}sica, Universidade Federal de Santa Catarina, 88040-900, Florian\'{o}polis-SC, Brazil}

\date{\today}

\begin{abstract}
We have studied ferromagnetic Permalloy clusters obtained by electrodeposition on n-type silicon. Magnetization measurements reveal hysteresis
loops almost independent on temperature and very similar in shape to those obtained in nanodisks with diameter bigger than 150nm. The spin 
configuration for the ground state, obtained by micromagnetic simulation, shows topological vortices with random chirality and polarization. 
This behavior in the small diameter clusters ($\sim$80nm), is attributed to the Dzyaloshinskii-Moriya interaction that arises in its hemispherical 
geometries. This magnetization behavior can be utilized to explain the magnetoresistance measured with magnetic field in plane and 
out of sample plane. 
\end{abstract}

\maketitle

  To lower the magnetostatic energy, ferromagnetic materials are arranged generally in magnetic domains separated by domain walls (DW), 
which are dislocated under external magnetic field in order to align the whole magnetization\cite{weiss}. In nanostructured ferromagnetic materials,
the magnetization fundamental state is strongly dependent on the geometry and other energetic configurations are oftentimes more likely, $e.g.$ 
nanodisks, which induce curvature of spin in plane with small displacement of spin from the border to the center, in order to keep the exchange 
interaction and cancel the dipolar energy. In the nanodisk center, the distance among spins becomes so small that the magnetization turns out of
the plane and this behavior makes the nanomagnet acts as a single giant spin\cite{domain}. When the disk thickness is much smaller then its diameter,
the magnetization aligns as a single domain in plane\cite{cowburn}. Its behavior is also important in nanostructured ellipsoidal monodomains,
which are often applied in different collective geometries, in systems knowed as artificial spin ice, where interesting phenomena arise 
as emergent magnetic monopoles\cite{mol}. The circular monodomain nanomagnets have been investigated in different systems, $e.g.$ vortex in 
nanodisks\cite{cowburn}, antidots samples\cite{araujo} or skyrmions\cite{skyrmion} that just differ from vortex with the spin in the borders turned out 
of the plane in direction opposite to the core polarization. Skyrmions frequently emerge in chiral materials under perpendicular external magnetic field with
Dzyaloshinskii-Moriya ($DM$) interaction \cite{dm,moryia}, and without $DM$ in systems with particular geometries\cite{sun}. The topological stability
in these structures, point them as promising in applications such as magnetic memory storage or logic devices\cite{mag-logic-storage,Iwasaki13}, 
with core and border polarization or chirality changed by external excitation such as $(AC)$ alternate current and magnetic field or spin polarized currents. In general, these structures
are fabricated by sophisticated techniques as Molecular Beam Epitaxy (MBE), $e_-$beam nanolitography or Focused Ion Beam (FIB).
\\
  In this work, we have utilized electrodeposition technique, which is less expensive, faster and more suitable in production lines than the techniques
cited above, to fabricate large area of Permalloy ($Py$) nanoclusters on silicon surfaces. We propose that the curvature of the hemispherical clusters shape is 
responsible for the arising of $DM$ interaction which allows the emergence of topological vortex excitations in very small structures.                          
\\
  The $Py$ clusters were obtained through galvanostatic deposition directly on the surface of $Si$ substrates. The substrates utilized were 
n-type (100) $Si$ samples with size of 1 $cm$ x 1 $cm$ cut from wafers commercially available with resistivity in the range of 1-10 $Ohm.cm$. 
Electrical contacts to each substrate for the electrodeposition were made through $GaIn$ back contact. An adhesive tape was used to mask off
all the substrate except for a circular area of 0.5 $cm^2$ on which the deposition was desired. Prior to deposition the substrates were immersed 
in a 5\% $HF$ solution for 5 to 10$s$, in order to remove oxide from the surface. The potentials were measured against a saturated calomel
electrode $(SCE)$, and a $Pt$ foil counter electrode was placed directly opposite the working electrode (substrate). $Py$ deposits on $Si$ 
were prepared from an aqueous electrolyte containing 30 mM $FeSO_4$, 700 mM $NiSO_4$, 20 mM $NiCl_2$, 16 mM saccharin, and 400 mM $H_3BO_3$, 
obtained from ref. \cite{elet1}, resulting in composition close to the $FeNi$ alloy (80 at.\% Ni and 20 at.\% Fe) for current density of 6.3 $mA/cm^2$ 
as determined previously \cite{elet2}. 
The electrodeposited samples were characterized by Scanning Electron Microscopy with Field Emission $(SEM-FEG)$ and Transmission Electron
Microscopy $(TEM)$. 
The magnetization behavior as a function of electrodeposition time was investigated by Superconducting Quantum Interference Device $(SQUID)$ and
the magnetoresistive measurements $(MR)$ were carried out using dc two-point probe method, where the two terminals were simple copper wires placed
on top of laterally prepared $GaIn$ ohmic contacts (2 x 4 mm) and 2 mm apart from each other, with magnetic field applied in plane and out of plane
in configuration transversal to the measuring current. The same configurations of $GaIn$ ohmic contacts were utilized in the current versus 
voltage $(I-V)$ measurements.        
\\
  In the beginning of the electrodeposition process, $Fe$ and $Ni$ solvated \'{i}ons in the electrolyte receive electrons from the substrate becoming adatoms. The adatoms 
migrate on the surface until finding a defect such as vacancies or kinks to nucleate. After the nucleation process, the clusters increase in size 
and the deposit evolves to compact thin film. Typical voltage transients were shown in Figure 1(a) for current density of 6.3 $mA/cm^2$. Only samples with superimposing
transients were considered for further measurements, in order to assure the required reproducibility of the properties under investigation.
Here we are interested in samples composed by isolated clusters on the surface that can be found by monitoring the 

\begin{figure}[!h]
		\center
		\includegraphics[width=8.5cm]{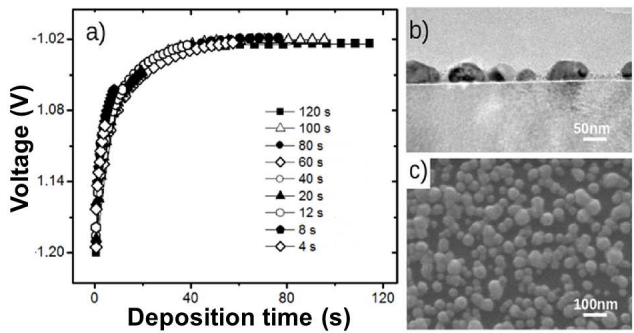}
		\caption{(a) Deposition transient measured against saturated calomel reference electrode. (b) Sample cross section measured by $TEM$ 
		and (c) Images of the clusters distribution obtained by $SEM-FEG$ .}
		\label{fig:fig1}
	\end{figure}	

electric percolation, which can be followed by electrical conductivity measurements, realized in the set of samples obtained in different
electrodeposition times. From the $(I-V)$ measurements, the time where the percolation occurs was found to be around 15$s$, so we focused our investigations in
samples with lower deposition times. From the micrographs performed by $SEM$ and $TEM$, presented in Figure 1(b) and 1(c), it is possible to note 
isolated clusters on the surface after 12$s$ of electrodeposition.
Hysteresis loops measured in the sample showed in Figure 1 for temperatures of 50K, 180K and 300K are shown in Figure 2.
\begin{figure}[!h]
		\center
		\includegraphics[width=8.0cm]{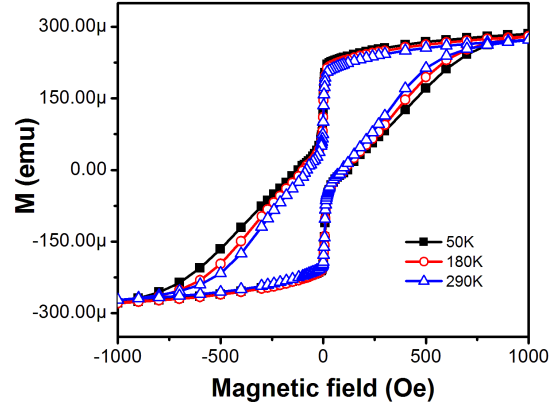}
		\caption{Magnetization measurements realized in SQUID for the sample with 12s of electrodeposition.}
		\label{fig:fig1}
	\end{figure}	
The similarity of the hysteresis measured in this work, with that observed in topological vortex states in nanodisks\cite{cowburn}, and its very low 
variation in the large temperature range investigated, are indications that each cluster bear a monodomain vortex excitation. In order to investigate
the spin dynamics in the clusters under external magnetic field, micromagnetic simulation were performed based on Landau-Lifshitz-Gilbert equation, 
\begin{equation}
\frac{\partial \vec{M}}{\partial t}= -\gamma \vec{M} \times \vec{H}_{eff} + \frac{\alpha}{M_{s}}\vec{M} \times \frac{\partial \vec{M}}{\partial t}\,,
\end{equation}
where $\gamma$ is the gyromagnetic ratio, $M_{s}$ the saturation magnetization and $H_{eff}$ is the effective magnetic field, which is composed
by external magnetic field, magneto-crystalline anisotropies, dipolar and exchange interactions. This equation is utilized to
determine the minimum energy and the transitions between spin configurations. For the iterations, we have utilized the software Object Oriented
MicroMagnetic Framework ($OOMMF$) \cite{oommf}, with parameters for $Py$ as $8.6$x$10^5 A/m$ for saturation magnetization, $13$x$10^-12 J/m^3$ 
for exchange constant and $0.5$ for damping coefficient.
\begin{figure*}[hbt]
		\center
		\includegraphics[width=15cm]{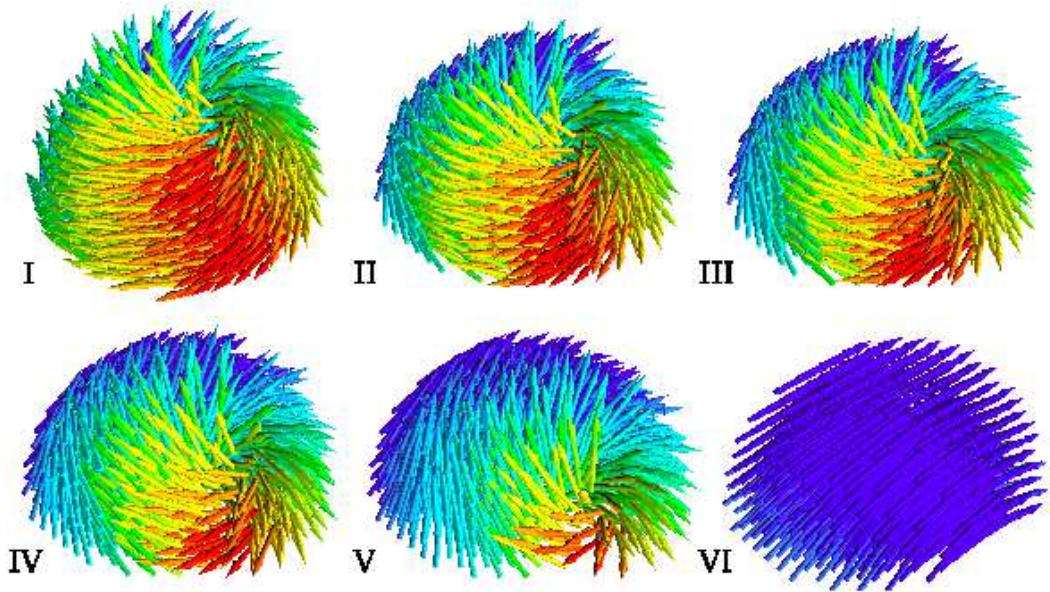}
		\caption{Spin configuration in the ground state I and the core delocation under external magnetic field II-V, until the saturated 
		monodomain following the external field orientation.}
		\label{fig:fig1}
	\end{figure*}	
The theoretical results were obtained in a model designed as hemispherical clusters of 80 nm 
diameter and 30 nm height, estimated from the experimental results. The separation of 80 nm among clusters in the model, was based on the average 
separation observed in the micrographs. In Figure 3I is presented the spin configuration obtained for the ground state in the simulated system and 
in Figures 3II-3VI the spin configuration under applied external magnetic field, with the vortex core been delocated to the borders until been 
excluded, resulting in a saturated monodomain aligned with the external magnetic field. During the decrease in external field, the vortex is created again in a cyclic process.
In Figure 4 are presented for comparison, the simulated hysteresis curve and the experimental one obtained at 50K. It is possible to note 
a similarity in shape between the curves, besides the deviation of about 10\% in absolute value of external field, which can be explained by the 
approximations in the model and to the fact that OOMMF simulation is performed at zero temperature. In the inset, it is presented the spin configurations 
for the ground state of a system composed by 4 elements, from this result it is possible to conclude that the ground state of magnetization in clusters 
array have curled vortex topology, with random chirality and polarizations of core vortices. 
In order to investigate the vortex formation on hemispherical clusters, let us considerate that magnetic materials (ferromagnetic or even 
antiferromagnetic) in two spatial dimensions may support topological excitations such as skyrmions and vortices. Vortices arise in classical 
magnetic systems containing an easy-plane anisotropy, which makes the spins prefer to point along the $XY$-plane. For instance, easy-plane 
ferromagnets are described by the Hamiltonian,
\begin{equation}
H=-J\sum_{i,j}[S_{i}^{x} S_{j}^{x}+ S_{i}^{y} S_{j}^{y}+ \lambda S_{i}^{z} S_{j}^{z}], 
\end{equation}
where $J>0$ is the exchange constant, $0\leq \lambda < 1$ the easy-plane anisotropy and $\vec{S}_{i}=(S_{i}^{x}, S_{i}^{y},S_{i}^{z})$ the classical
spin vector at site $i$. Considering the most realistic discrete lattice case, and depending on the range of $\lambda$, such an easy-plane system
supports two types of static vortices: the in-plane vortex (in which all spins are confined to the $XY$-plane\cite{???}) and the 
out-of-plane vortex (in which some spins around the vortex center can point perpendicularly to the $XY$-plane). Indeed, considering a critical value
of $\lambda$, denoted by $\lambda_{c}$, then for the range $\lambda <\lambda_{c}$, the stable excitation is the in-plane vortex, while for 
$\lambda >\lambda_{c}$, the out-of-plane vortex becomes stable \cite{Wysin94}. The stability of these solutions has only been determined via computer
simulations. The critical anisotropy $\lambda_{c}$ depends on the lattice geometry: for the square lattice, $ \lambda_{c}=0.72$; similarly 
$\lambda_{c}=0.86$ for the hexagonal lattice and $\lambda_{c}=0.62$ for the triangular lattice \cite{Wysin94}. Qualitatively, similar results can be
obtained for $2D$ easy-plane antiferromagnetic systems. 
Broken mirror symmetry at surfaces/interfaces of magnetic nanostructures, induces chiral DM interactions which may strongly affect the magnetic properties
of the system, for example allowing the possibility of vortex with a well defined sense of rotation (curl vortex with a chiral sense). 
\begin{figure}[!h]
		\center
		\includegraphics[width=8.0cm]{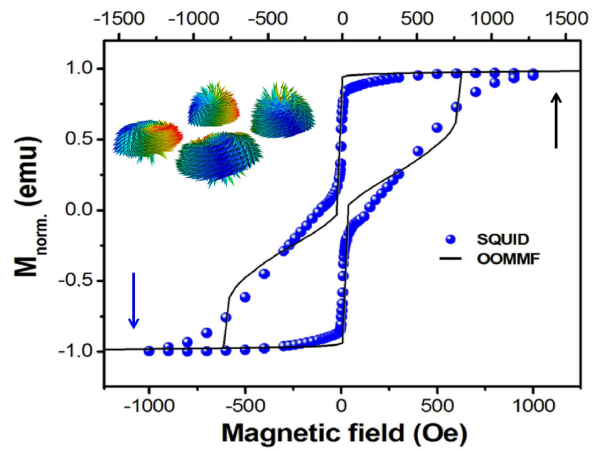}
		\caption{Hysteresis loop obtained with the micromagnetic simulation (line), compared with the experimental measurement at 50K (symbol).}
		\label{fig:fig1}
	\end{figure}	
When the manifold that supports the spins is a curve one, or when the spin remains in a curved manifold, the curvature results in two additional effective interactions, originated from the exchange interactions\cite{kravchuck}, one is analogue to magnetic
anisotropy and another to the DM interaction. 
The equilibrium state of the magnetic hemispheres, can be understood as a competition between these
effective curvature induced interactions. For example if the exchange energy density is wroten in function of curvilinear coordinates, the terms are
the isotropic exchange energy, anisotropy energy that depends on the geometry of the manifold, and the effective DM interaction. The last one can 
explain the vortex formation on a hemispherical cluster. When there is only exchange interaction (without DM interaction), the competition between the
in-plane exchange energy, tending to extend the vortex core, and the uniaxial anisotropy, favoring its shrinking, determines the equilibrium size of
the core. It may occupy the whole network being of any size, but in magnetic nanodisks, for example, where vortices are naturally the ground state of
these systems and they can be directly observed by experimental techniques\cite{Miltat02,Guslienko06}, the typical size of the vortex is down to $150$ nm
\cite{sun}.
The DM interaction, extends sizes of the vortices with favorable rotation sense and compresses the vortices with opposite chirality \cite{butenko}.
The ratio between the DM constant interaction $D_0$ and the exchange interaction $J$, determines the size of the vortex, being the vortex stable in
a range of values ​​where the competition between the energy is favorable to the emergence of the vortex, minimizing the energy of the system.
\begin{figure}[!h]
		\center
		\includegraphics[width=8.0cm]{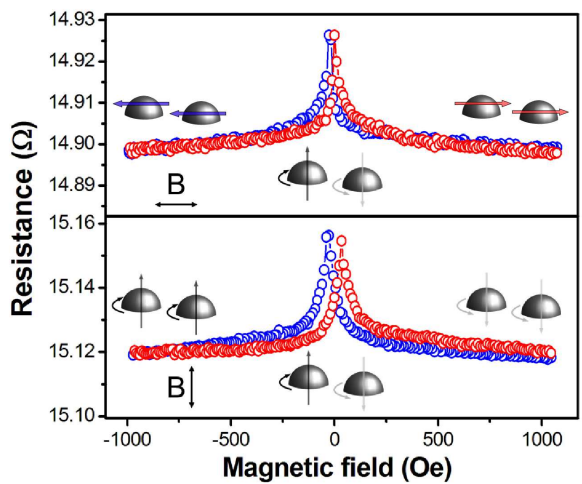}
		\caption{Magnetoresistance measured with external field in plane and out of sample plane. The cartoons show the spin configuration in 
		each part of the curves.}
		\label{fig:fig1}
	\end{figure}	
The greater the ratio $D_0/J$, smaller the size of the vortex. The DM interaction induced by curvature, has a coupling constant $D_0$ that is a 
function of the geometry manifold where the spins reside.
To a hemispherical one, $D_0$ can be divided in two parts, which are functions of $1/r$
and $1/r^3$, being $r$ the radius of the hemisphere. Then to small values of $r$, more larger is the DM coupling and smaller are the size of the vortex.
The experimental detection of this vortex, can be used to asses the curvature induced DM interaction and measure the strength of the induced DM coupling
in this system.\\
In Figure 5 are presented the magnetoresistive measurements performed in this work. Due to its magnetic behavior, the clusters array presents isotropic 
magnetoresistance\cite{apl2008, balestar}, similar to the Giant Magnetorestance (GMR) effect in ferromagnetic-nonmagnetic multilayers\cite {fert, grumbert}.
The ferromagnetic configurations obtained with the monodomains aligned to the external magnetic field, in sample plane, and in the every core polarization
aligned to the external magnetic field, applied out of the sample plane, provide lower resistive path to the spin polarized current which must flow 
throughout the silicon. The higher resistance close to the zero field, occurs due to the antiferromagnetic configuration of the curling vortex states 
with opposite chirality and core polarizations, in order to diminish magnetostatic energy.
\\
  In summary we have investigated the behavior of magnetization in Py clusters electrodeposited on silicon. From the experimental hysteresis and 
spin dynamics, realized by micromagnetic simulation, we have observed topological vortex configuration in the clusters ground state, which was 
attributed to the emergence of DM interactions in its hemispherical geometry. The vortex core alignment under out of plane external magnetic field
or monodomains formation under in plane external magnetic field, enables the investigation of such systems in magnetoresistive devices.    
\\
  The authors are grateful to Daisy de Melo Gomes (in memoriam) for the TEM images. They also thank CAPES, CNPq, FAPESC and FAPEMIG (Brazilian agencies)
for partial financial support.




\thebibliography{99}
    
    \bibitem{weiss}P. Weiss, J. Phys. \textbf{6}, 661 (1907)

    \bibitem{domain}A. Hubert and H. Schafer, Springer Berlin (1998).
    
    \bibitem{cowburn}R. P. Cowburn, D. K. Koltsov, A. O. Adeyeye, M. E. Welland and D. M. Tricker, Phys. Rev. Lett. \textbf{83}, 1042 (1999).

    \bibitem{mol} L. A. M\'{o}l, R. L. Silva, R. C. Silva, A. R. Pereira, W. A. Moura-Melo and B. V. Costa, J. Appl. Phys. \textbf{106}, 063913 (2009). 

    \bibitem{araujo}C. I. L. de Araujo, R. C. Silva, I. R. B. Ribeiro, F. S. Nascimento, J. F. Felix, S. O. Ferreira, L. A. S. M\'{o}l, W. A. Moura-Melo,
    A. R. Pereira, Appl. Phys. Lett. \textbf{104}, 092402 (2014).
			
    \bibitem{skyrmion} R. L. Silva, L. D. Secchin, W. A. Moura-Melo, A. R. Pereira, R. L. Stamps, Physical Review B \textbf{89} p. 054434-054434-8 (2014). 

    \bibitem{dm}I. Dzyaloshinsky, J. Phys. Chem. Solids \textbf{4}, 241 (1958).

    \bibitem{moryia}T. Moriya, Phys. Rev. \textbf{120}, 91 (1960).
.    
    \bibitem{sun} L. Sun,R. X. Cao, B. F. Miao, Z. Feng, B. You, D. Wu, W. Zhang, A. Hu and H.F. Ding, Phys. Rev. Lett. \textbf{110}, 167201 (2013).

    \bibitem{mag-logic-storage}M. Rahm, J. Stah and D. Weiss, Appl. Phys. Lett. \textbf{87}, 182107 (2005).
    
    \bibitem{Iwasaki13}J. Iwasaki, M. Mochizuki and N. Nagaosa, Nature Nanotech. \textbf{8}, 742 (2013). 
        
    \bibitem{elet1}J. -M. Quemper, S. Nicolas, J. P. Gilles, J. P. Grandchamp, A. Bosseboeuf, T. Bourouina and E. Dufour-Gergam, Sens. Actuators A
    \textbf{74} (1-3) pp. 1-4 (1999).

    \bibitem{elet2}E. R. Spada, L. S. de Oliveira, A. S. da Rocha, A. A. Pasa, G. Zangari and M. L. Sartorelli, J. Magn. Magn. Mater. \textbf{272}, e891 (2004).

    \bibitem{oommf}M. J. Donahue and D. G. Porter, OOMMF Object Oriented MicroMagnetic Framework, software $NIST$ v1.2a3 (2004).
    
    \bibitem{???}M. Getzlaff, Springer Berlin (2008).
  
    \bibitem{Wysin94}G. M. Wysin, Phys. Rev. B \textbf{49}, 8780 (1994).

    \bibitem{kravchuck}Y. Gaididei, V. P. Kravchuk and D. D. Sheka, Phys. Rev. Lett. \textbf{112}, 257203 (2014).

    \bibitem{Miltat02} J. Miltat and A. Thiaville, Science \textbf{298}, 555 (2002). 

    \bibitem{Guslienko06}K. Y. Guslienko, X. F. Han, D. J. Keavney, R. Divan and S. D. Bader, Phys. Rev. Lett. \textbf{96}, 067205 (2006).

    \bibitem{butenko}A. B. Butenko, A. A. Leonov, U. K. Rossler and A. N. Bogdanov, Phys. Rev. B \textbf{82}, 052403 (2010).
    
    \bibitem{apl2008}C. I. L. de Araujo, M. L. Munford, R. G. Delatorre, A. V. Da Cas, V. C. Zoldan, A. A. Pasa, N. G. Garcia, Appl. Phys. Lett. \textbf{92}, 222101 (2008).

    \bibitem{balestar}A. Ballestar, C. I. L. de Araujo, R. G. Delatorre, A. A. Pasa, N. G. Garcia, J Supercond Nov Magn \textbf{22}, p. 737-748 (2009).
     
    \bibitem{fert}M. N. Baibich, J. M. Broto, A. Fert, F. Nguyen Van Dau, F. Petroff, P. Etienne, G. Creuzet, A. Friederich and J. Chazelas, 
    Phys. Rev. Lett. \textbf{61}, 2472 (1988).  
     
    \bibitem{grumbert}G. Binasch, P. Gr\"{u}nberg, F. Saurenbach and W. Zinn, Phys. Rev. B \textbf{39}, 4828(R) (1989)
\end{document}